\documentclass[12pt]{article}
\usepackage[utf8]{inputenc}
\usepackage{graphicx}
\usepackage{geometry}
\usepackage{caption}
\usepackage{float}
\usepackage{amsmath}
\usepackage{authblk} 

\title{Comparative Simulation of CEvNS Recoil Observables in Geant4 Using  Germanium, Argon, and Xenon Targets}
\author{Yusuf Havvat}
\affil{Department of Physics, Çukurova University, Adana, Turkey\\
\texttt{havvatyusuf249@gmail.com}}
\date{\today}

\begin{document}

\maketitle
\tableofcontents
\newpage

\section{Abstract}
Coherent Elastic Neutrino-Nucleus Scattering (CEvNS) is a low-energy process in which a neutrino interacts coherently with an entire nucleus, leading to a detectable nuclear recoil. This study presents a detailed Geant4-based simulation of CEvNS events for three commonly used target materials in neutrino detection: Germanium, Argon, and Xenon. A custom CEvNS process was developed and integrated into the Geant4 framework, incorporating realistic differential cross-section models and the Helm form factor. ROOT-based analysis was used to evaluate recoil energy spectra, angular distributions, and form factor suppression effects for each target. The results confirm expected physical trends: heavier nuclei like Xenon produce smaller recoil energies and exhibit stronger form factor suppression compared to lighter nuclei. This work demonstrates a complete end-to-end simulation of CEvNS from particle generation to recoil analysis and offers a reproducible framework that can support both theoretical modeling and experimental interpretation in neutrino physics and dark matter search experiments

\section{Introduction}
Coherent elastic neutrino-nucleus scattering (CEvNS) is a fundamental process predicted by the Standard Model, in which a low-energy neutrino scatters elastically off an entire nucleus via neutral-current interactions. First observed by the COHERENT Collaboration, CEvNS provides a powerful probe of electroweak physics, nuclear structure, and potential new physics beyond the Standard Model.

A precise understanding of CEvNS is also crucial for the design and optimization of neutrino detectors. The process is particularly relevant for experiments employing heavy noble gases and semiconductors, such as xenon, argon, and germanium detectors. These targets differ in nuclear recoil energy spectrum, form factor suppression, and detection thresholds, making comparative studies essential.

In this work, we present a detailed simulation of CEvNS interactions in Ge, Ar, and Xe targets using the Geant4 toolkit. The simulations incorporate realistic nuclear form factors and detector response conditions. The results are analyzed with ROOT to extract distributions of recoil energy, scattering angle, and time structure. Our aim is to highlight systematic differences between light and heavy targets, with implications for
future experimental searches.

\section{Methodology}
\label{sec:methods}

\subsection{Overview}
We simulate Coherent Elastic Neutrino--Nucleus Scattering (CEvNS) for three detector targets
(Germanium, Argon, Xenon) using the \textsc{Geant4} simulation toolkit~\cite{Agostinelli2003,Allison2016}. 
A user-defined process (\texttt{CEvNSProcess}) is implemented to sample nuclear recoils according 
to the Standard Model differential cross section, incorporating a nuclear form factor. 
All simulated data are analyzed using the \textsc{ROOT} framework~\cite{Brun1997}.

\subsection{Neutrino source and kinematics}
Unless stated otherwise, we assume an electron antineutrino spectrum representative of a
reactor-like source with $E_\nu \in [0,10]~\mathrm{MeV}$.\footnote{A Gaussian or
piecewise-analytic parameterization may be used for fast sampling; the exact spectrum can
be replaced by an SNS-like spectrum in a sensitivity study.} Neutrinos are generated via
\texttt{PrimaryGeneratorAction} with momentum along $+\hat{z}$ toward the detector center.
For an elastic CEvNS interaction on a nucleus of mass $m_N$, the recoil energy is
$T\equiv E_\mathrm{recoil}$ and the four-momentum transfer satisfies $q^2=2m_NT$ (non-relativistic nucleus).

\subsection{CEvNS differential cross section}
The Standard Model CEvNS differential cross section is

\begin{equation}
\frac{d\sigma}{dT}(E_\nu,T) \;=\;
\frac{G_F^2\, m_N}{4\pi}\; Q_w^2\;
\left(1 - \frac{m_N T}{2 E_\nu^2}\right)\; F^2(q^2),
\quad q^2 = 2 m_N T,
\label{eq:dsigmadT}
\end{equation}

where $G_F$ is the Fermi constant and 
$Q_w = N - \bigl(1 - 4\sin^2\theta_W\bigr)Z \simeq N$ 
is the weak nuclear charge (with $Z$ protons and $N$ neutrons). 
Equation~\eqref{eq:dsigmadT} is sampled to generate recoil energies $T$ 
and corresponding scattering angles $\theta$ using two-body kinematics 
for a given neutrino energy $E_\nu$. 
This ensures energy–momentum conservation for each simulated CEvNS event.

The sampling method employs the differential cross-section weight 
$d\sigma/dT(E_\nu, T)$ combined with the nuclear form factor $F^2(q^2)$ 
to account for coherence loss at higher momentum transfers. 
Recoil events are generated using a Monte Carlo acceptance–rejection technique 
to preserve the true physical distribution of scattering probabilities.

\subsection{Nuclear form factor (Helm)}
To account for the loss of coherence at finite momentum transfer, 
the Helm form factor~\cite{Helm1956} is employed:

\begin{equation}
F(q) = 3 \, \frac{j_1(qR_0)}{qR_0} \, \exp\!\left[-\frac{(qs)^2}{2}\right],
\label{eq:helm}
\end{equation}

where $j_1$ is the spherical Bessel function, $R_0$ an effective nuclear radius, 
and $s$ the surface thickness. Unless otherwise specified, we take 
$R_0 = 1.2\,A^{1/3}\,\mathrm{fm}$ and $s = 0.9\,\mathrm{fm}$, 
a common choice in CEvNS phenomenology. We tabulate target parameters used in the simulation:

\begin{center}
\begin{tabular}{lcccc}
\hline
Target & $Z$ & $N$ & $R_0$ [fm] & $s$ [fm] \\
\hline
Ge & 32 & 41 & $1.2\,A^{1/3}$ & 0.9 \\
Ar & 18 & 22 & $1.2\,A^{1/3}$ & 0.9 \\
Xe & 54 & 77 & $1.2\,A^{1/3}$ & 0.9 \\
\hline
\end{tabular}
\end{center}

If experimentally calibrated $R_0$ or $s$ values are available, they can be 
substituted in Eq.~\eqref{eq:helm} and the table.

\subsection{Detector geometry and materials}
The detector is a cubic active volume fully filled with the target material, centered in a
vacuum world volume whose dimensions are chosen to suppress boundary effects. Targets
use \textsc{Geant4} materials \texttt{G4\_Ge}, \texttt{G4\_Ar}, and \texttt{G4\_Xe}
with natural isotopic abundances.

\subsection{Physics list and custom process}
We build upon \texttt{G4VModularPhysicsList} and include
\texttt{G4EmStandardPhysics} (for standard electromagnetic interactions),
\texttt{G4DecayPhysics} (for unstable particle decays),
and our user-defined \texttt{CEvNSProcess}. 
The latter is registered for the electron neutrino (\texttt{G4NeutrinoE}) 
within the \texttt{ConstructProcess()} method to ensure that 
the coherent elastic neutrino–nucleus scattering interaction is 
invoked whenever a neutrino passes through the detector volume. 

The process is attached to the particle definition using 
the \texttt{G4ProcessManager} interface:
\texttt{pmanager->AddDiscreteProcess(new CEvNSProcess());}.
This guarantees that the CEvNS interaction is treated as a 
discrete nuclear scattering event rather than a continuous process. 
Each event is independently evaluated to determine whether 
a scattering occurs based on the differential cross section 
defined in Eq.~\eqref{eq:dsigmadT}, incorporating the nuclear form factor
from Eq.~\eqref{eq:helm}.

\subsection{Simulation workflow, statistics, and outputs}
Each event produces at most one CEvNS recoil. For each target we simulate
$\mathcal{O}(10^4$--$10^5)$ primaries to reduce statistical fluctuations in the low-$T$ tail.
At end-of-event we record: recoil energy $T$ (keV), scattering angle $\theta$ (rad),
time stamp $t$ (ns), and $F(q^2)$. Histograms and density plots (TH1/TH2) are saved as
PDFs and the full TTree is persisted for post-processing. We report histogram means
and standard deviations and propagate Poisson uncertainties on binned spectra.

\subsection{Units and conventions}
We use \textsc{Geant4} internal units and convert to keV (for $T$), MeV (for $E_\nu$),
and ns (for $t$) in output. Axis labels and color bars explicitly include units.
Angles are in radians unless noted otherwise.

\subsection{Detector Information}
In the simulation environment, the detector is defined as a cubic volume. The detector volume is filled with the target material to be analyzed. Germanium (Ge), Argon (Ar), and Xenon (Xe) nuclei were defined separately, and simulations were run independently for each target material. The detector's exterior is surrounded by a vacuum, and the outer volume is defined as the "world volume." The world volume was chosen to be large compared to the detector to prevent edge effects.

\subsection{Particle Production and Energy Spectrum}
Neutrino-like particles were created in the simulation to trigger the CEvNS process. These particles:
The electron-type neutrino is defined by its character (G4NeutrinoE). It is designed to have an energy spectrum that is close to constant or Gaussian, extending up to 10 MeV. The direction of momentum is defined as being along the z-axis and is directed towards the center of the detector. This process is configured with the PrimaryGeneratorAction class, and a single neutrino production is provided for each event.
\subsection{Definition of the CEvNS Process}

Since the Coherent Elastic Neutrino–Nucleus Scattering (CEvNS) process is not included in the standard 
\textsc{Geant4} physics libraries, a dedicated user-defined class named \texttt{CEvNSProcess} was developed 
for this study. Rather than serving as a simple example, this class provides a physically consistent model 
of low-energy neutrino–nucleus interactions, capable of reproducing recoil observables across different target materials.

The process is derived from \texttt{G4VDiscreteProcess} and is explicitly applicable only to electron neutrinos. 
A fixed mean free path of $50~\mathrm{cm}$ was introduced to control the event generation rate and ensure sufficient 
statistics during simulation. For each event, the incoming neutrino energy $E_\nu$ and the target nuclear mass $M$ 
are used to calculate the recoil energy $T$ through two-body elastic scattering kinematics:
\[
T = \frac{2E_\nu^2(1-\cos\theta)}{M + 2E_\nu(1-\cos\theta)}.
\]
The momentum transfer is defined as $q^2 = 2MT$, and the probability of the scattering event is weighted 
by the nuclear form factor $F(q^2)$, implemented according to the Helm parameterization:
\[
F(q) = 3\,\frac{j_1(qR_0)}{qR_0}\,\exp\!\left[-\frac{(qs)^2}{2}\right],
\]
where $R_0 = 1.2A^{1/3}$~fm and $s = 0.9$~fm represent the effective nuclear radius and surface thickness, respectively. 
This formulation captures the loss of coherence at higher momentum transfer and ensures that heavy nuclei 
exhibit realistic suppression of the CEvNS cross section.

During the simulation, a random scattering angle $\theta$ is sampled from a uniform distribution, 
and the resulting recoil direction is generated accordingly. For each valid scattering event, a recoiling ion 
(\texttt{G4DynamicParticle}) is created and added to the simulation with its energy and direction defined by 
the calculated recoil parameters. The process records the following quantities for each event: 
event ID, target atomic and mass numbers ($Z$, $A$), recoil energy $E_\mathrm{recoil}$, 
scattering angles ($\theta$, $\phi$), position, global time, and the applied form factor $F(q^2)$. 
All quantities are saved in \textsc{ROOT} format through the \texttt{RootManager} class for post-simulation analysis.

Unlike standard weak interaction models in \textsc{Geant4}, this custom process explicitly incorporates 
the Helm nuclear form factor and a physics-based weighting of scattering probabilities. 
Consequently, \texttt{CEvNSProcess} serves not only as a dedicated CEvNS model but also as a flexible 
simulation framework that allows comparative studies among different nuclear targets 
and provides predictive guidance for future detector materials.
 It operates with a single core per event. Includes differential cross-section information (optionally integrated with the Helm form factor). The Helm model was used as the form factor, and definitions were made that considered the nuclear structure of the cores. The momentum transfer q was used to calculate the form factor F(q²), which was then integrated to affect the scattering probability.
\subsection{Physics List}
The physics list is based on the modular Geant4 framework~\cite{Allison2016}.
G4EmStandardPhysics: Electromagnetic processes
G4DecayPhysics: Unstable particle decays
CEvNSProcess: Coherent Elastic Neutrino-Nucleus Scattering process
Thanks to this structure, the CEvNS process was carried out in a controlled manner alongside other physical processes.

\subsection{Data Recording and Analysis}
At the end of each event, information such as the kinetic energy of the recoiling nucleus,
the time of interaction, position and momentum direction (scattering angle $\theta$),
and the applied nuclear form factor $F(q^2)$ are recorded through the \texttt{RootManager} class.
All quantities are stored within the \texttt{RecoTree} data structure in a ROOT file (\texttt{output.root}).

Subsequent analysis is performed using \textsc{ROOT} macros and scripts.
Each observable is represented using ROOT histogram classes such as 
\texttt{TH1F} (1D distributions) and \texttt{TH2F} (2D correlations), 
and displayed on a \texttt{TCanvas} object.
Each histogram and correlation plot is saved using the 
\texttt{SaveAs()} function, exporting the figures in PDF format 
(e.g., \texttt{TCanvas::SaveAs("energy\_Ge.pdf")}).

All axes are labeled with physical units, and figure captions explicitly 
state the target material and observable.
Statistical uncertainties are evaluated from bin counts assuming 
Poisson fluctuations, and reported as $\sigma = \sqrt{N}$.
The mean and standard deviation for each observable are 
automatically computed and included in the analysis summary.

\subsection{Recoil Energy Distributions}

\begin{figure}[H]
    \centering
    % Germanium
    \begin{minipage}{0.45\textwidth}
        \centering
        \includegraphics[width=\linewidth]{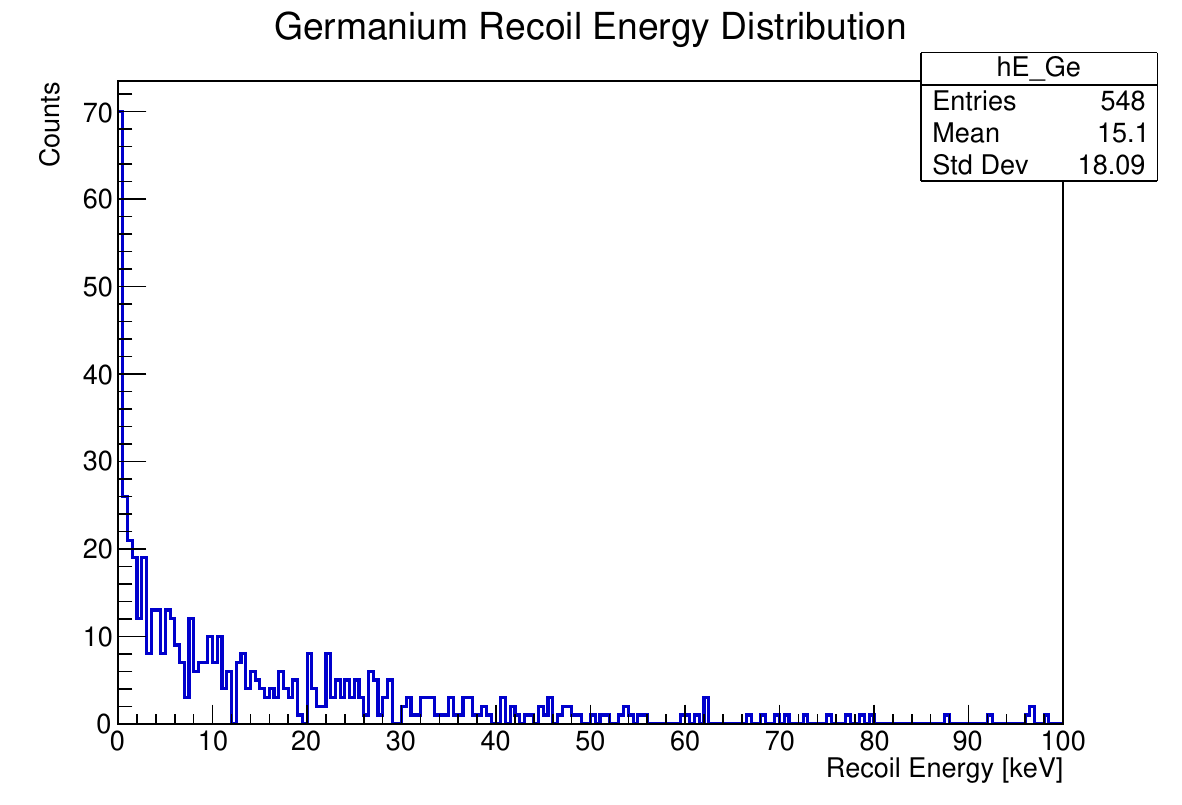}
        \caption*{Germanium}
    \end{minipage}
    % Argon
    \begin{minipage}{0.45\textwidth}
        \centering
        \includegraphics[width=\linewidth]{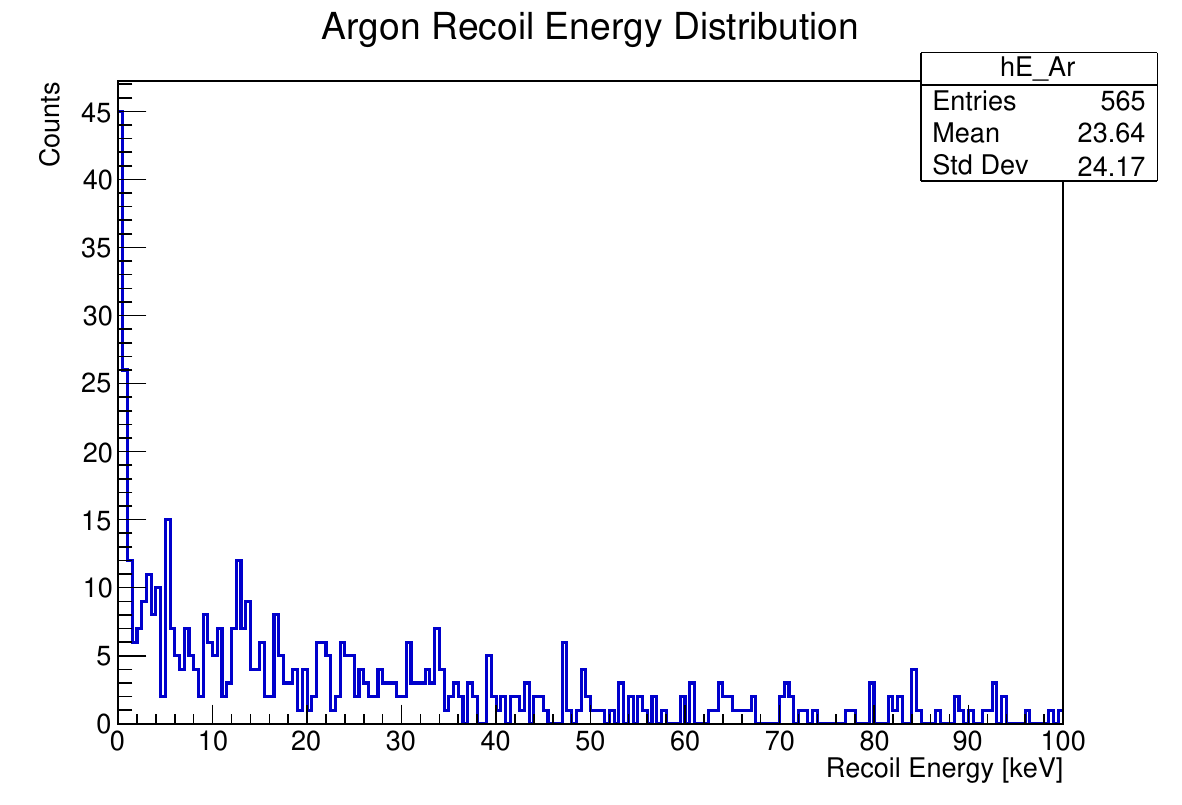}
        \caption*{Argon}
    \end{minipage}

    \vspace{0.3cm} % Satırlar arası boşluk

    % Xenon
    \begin{minipage}{0.6\textwidth}
        \centering
        \includegraphics[width=\linewidth]{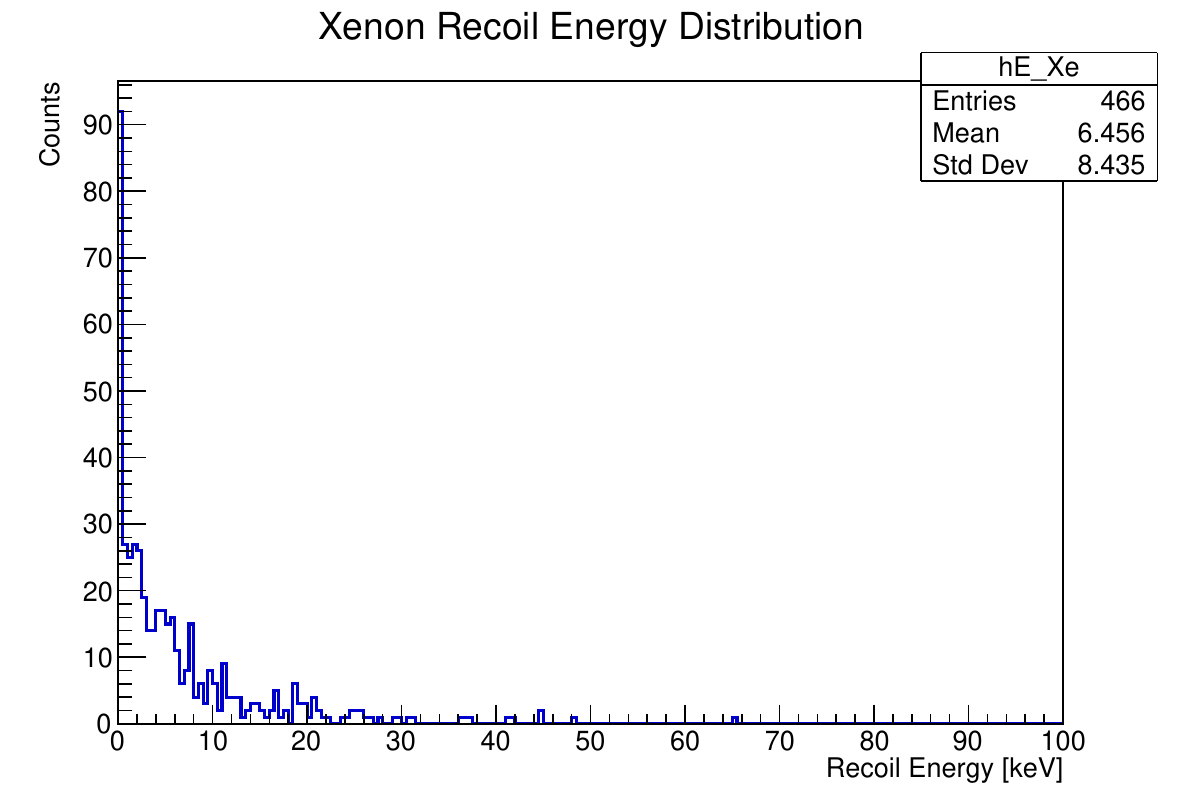}
        \caption*{Xenon}
    \end{minipage}

    \caption{Recoil energy spectra for Germanium, Argon, and Xenon targets.}
    \label{fig:energy_comparison}
\end{figure}

The recoil energy spectra for Germanium, Argon, and Xenon targets are shown in 
Fig.~\ref{fig:energy_comparison}. A clear dependence on nuclear mass is observed:

\begin{itemize}
    \item \textbf{Argon ($\langle T \rangle \approx 23.6$ keV):} 
    As the lightest target nucleus considered, argon produces the highest recoil energies. 
    The distribution extends up to $\sim$100 keV, with a broad spread (Std. Dev. $\approx$ 24 keV). 
    This makes argon recoils easier to detect with relatively modest detector thresholds, 
    consistent with observations from the COHERENT experiment.... consistent with the COHERENT measurement on argon~\cite{COHERENT2017}.    
    
    \item \textbf{Germanium ($\langle T \rangle \approx 15.1$ keV):} 
    Germanium lies in an intermediate regime. Recoil energies are lower than argon but 
    higher than xenon, with a moderate spread (Std. Dev. $\approx$ 18 keV). 
    This balance, combined with the availability of ultra-low-noise HPGe detectors, 
    makes germanium an attractive choice for CEvNS studies and has enabled 
    recent successful measurements (e.g., CONUS, CONNIE).... in agreement with germanium-based CEvNS studies~\cite{COHERENT2017,CONUS2021}.
    
    \item \textbf{Xenon ($\langle T \rangle \approx 6.5$ keV):} 
    Owing to its large nuclear mass, xenon produces the lowest recoil energies, 
    predominantly below 20 keV. The narrow distribution (Std. Dev. $\approx$ 8.4 keV) 
    implies that only detectors with thresholds below $\sim$5 keV can effectively 
    measure CEvNS in xenon, as highlighted in XENONnT and PandaX-4T analyses.
\end{itemize}... consistent with recent XENONnT and PandaX-4T analyses~\cite{XENONnT2023,PandaX2021}.

\noindent
These results confirm the expected scaling of recoil energy with target mass: 
lighter nuclei (argon) provide higher recoil energies but lower event rates, while 
heavier nuclei (xenon) yield lower recoil energies but higher total cross sections. 
Germanium represents a compromise, offering manageable thresholds and mature 
detector technologies. This comparison underscores the importance of 
threshold optimization in CEvNS detector design.

\subsection{Time–Energy Correlations}

\begin{figure}[H]
    \centering
    % Germanium
    \begin{minipage}{0.45\textwidth}
        \centering
        \includegraphics[width=\linewidth]{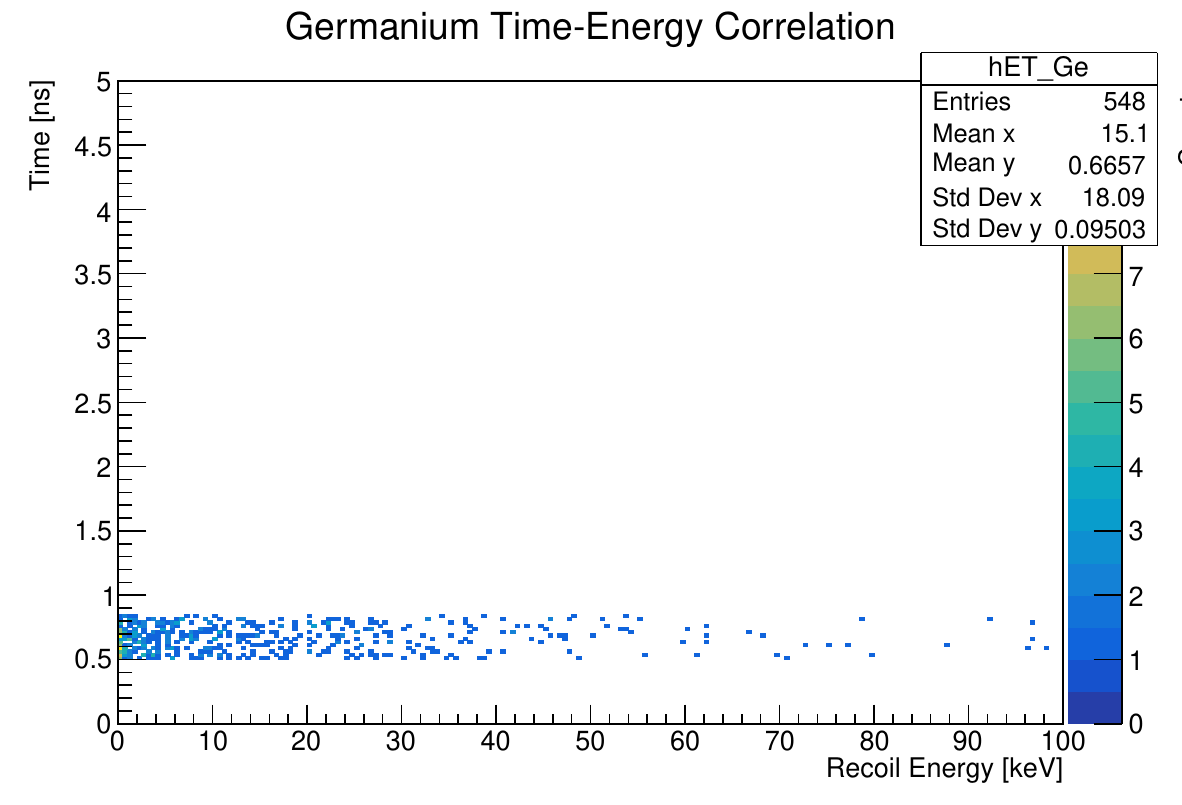}
        \caption*{Germanium}
    \end{minipage}
    % Argon
    \begin{minipage}{0.45\textwidth}
        \centering
        \includegraphics[width=\linewidth]{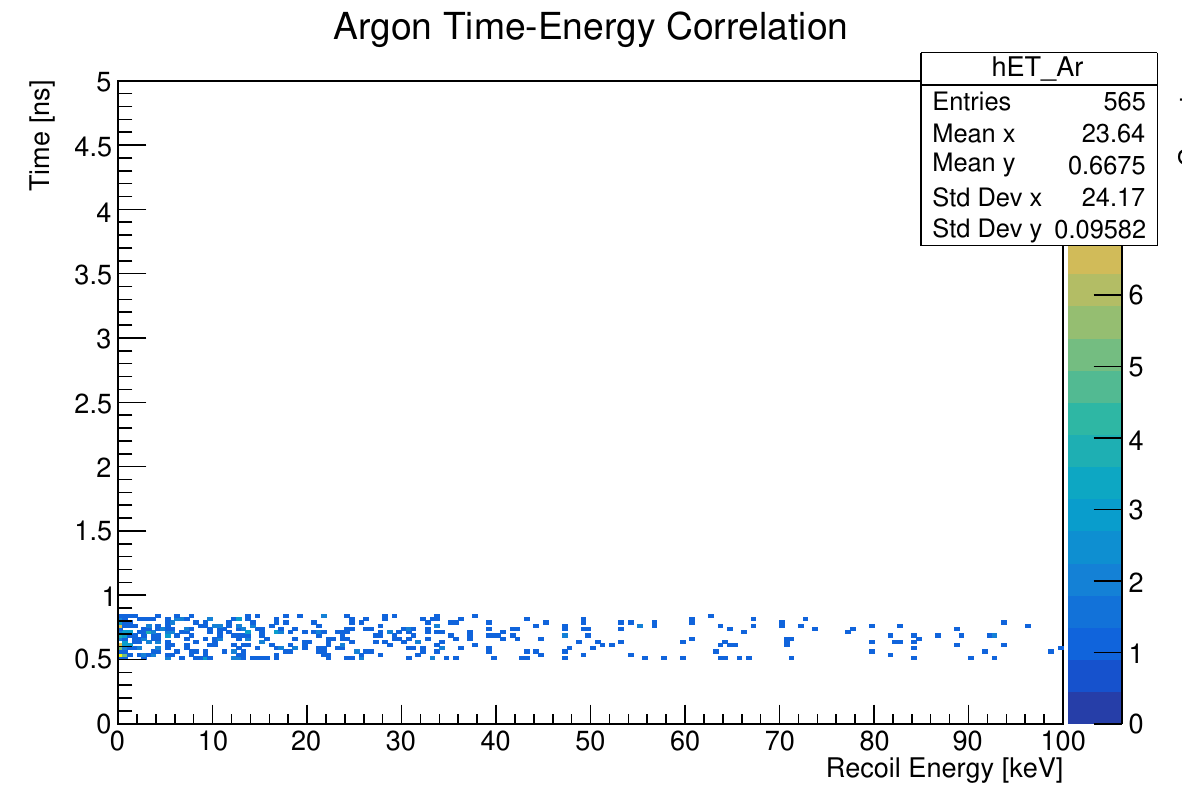}
        \caption*{Argon}
    \end{minipage}

    \vspace{0.3cm} % Satırlar arası boşluk

    % Xenon
    \begin{minipage}{0.6\textwidth}
        \centering
        \includegraphics[width=\linewidth]{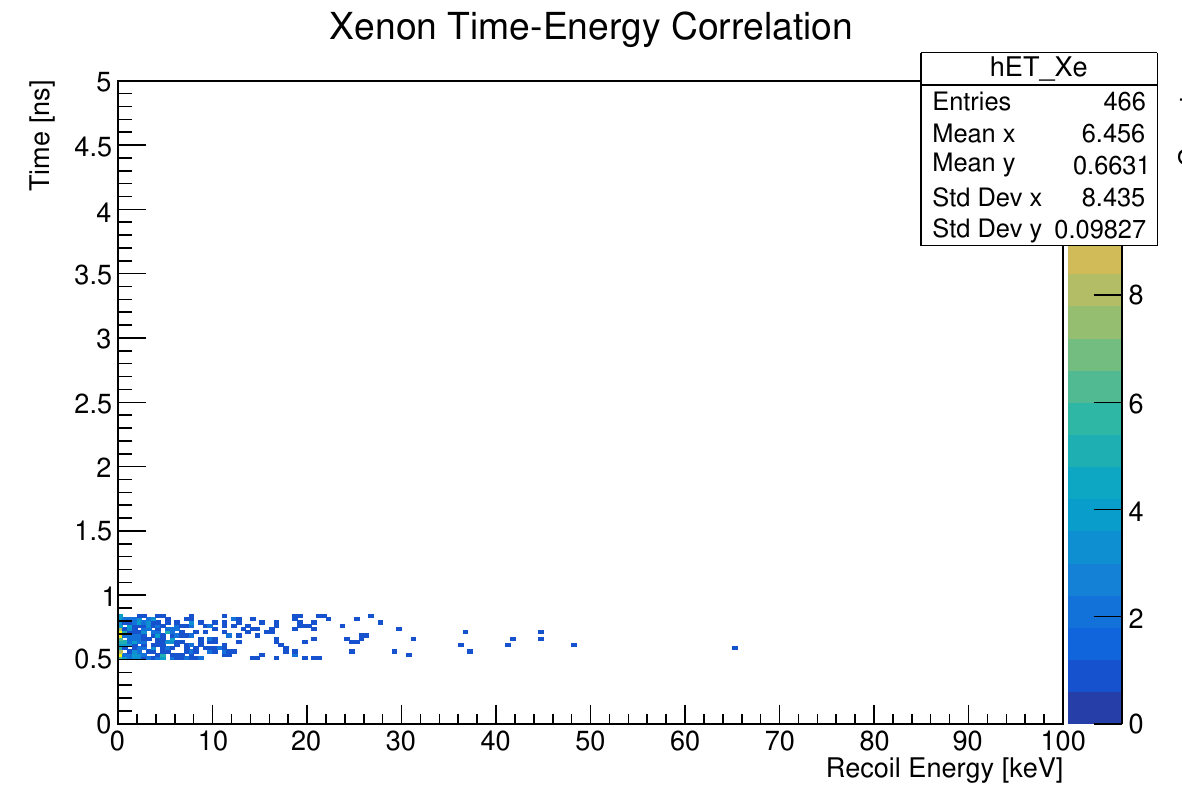}
        \caption*{Xenon}
    \end{minipage}

    \caption{Comparison of time–energy correlations for Germanium, Argon, and Xenon targets. 
    The mean event time is approximately $0.66$~ns in all cases, with small statistical variations 
    ($\sigma_t \approx 0.095$~ns). No significant dependence of recoil energy on event timing is observed, 
    confirming the stability of CEvNS interactions within the simulated neutrino flux window. 
    These results are consistent with steady-state neutrino sources such as COHERENT and reactor-based CEvNS experiments~\cite{COHERENT2017,CONUS2021}.}
    \label{fig:timeenergy_comparison}
\end{figure}

\noindent
\textbf{Argon ($\langle E \rangle \approx 23.6~\mathrm{keV}$, $\langle t \rangle \approx 0.668~\mathrm{ns}$):}  
The highest recoil energies are reached, but the time distribution remains flat, 
indicating that CEvNS events in argon occur randomly within the neutrino flux window. 
This uniformity is characteristic of continuous reactor-like sources~\cite{COHERENT2017}.  

\textbf{Germanium ($\langle E \rangle \approx 15.1~\mathrm{keV}$, $\langle t \rangle \approx 0.666~\mathrm{ns}$):}  
An intermediate recoil spectrum with a similarly stable time profile is observed, 
showing that timing information in germanium detectors is not strongly energy-dependent.  

\textbf{Xenon ($\langle E \rangle \approx 6.5~\mathrm{keV}$, $\langle t \rangle \approx 0.663~\mathrm{ns}$):}  
Although xenon recoils have the lowest energies, the time spread is comparable to argon and germanium, 
reflecting the smooth structure of the simulated neutrino beam and the coherence of CEvNS timing behavior.  

\noindent
Overall, these results confirm that for steady or reactor-like neutrino sources, 
the detector timing resolution need not depend on recoil energy. 
The flat time–energy correlations across different targets validate the assumption 
of a uniform neutrino arrival profile and support the feasibility of time-tagged CEvNS measurements 
in large-scale experiments~\cite{COHERENT2017,CONUS2021}.

\subsection{Scattering Angle Distributions}

\begin{figure}[H]
    \centering
    % Germanium
    \begin{minipage}{0.45\textwidth}
        \centering
        \includegraphics[width=\linewidth]{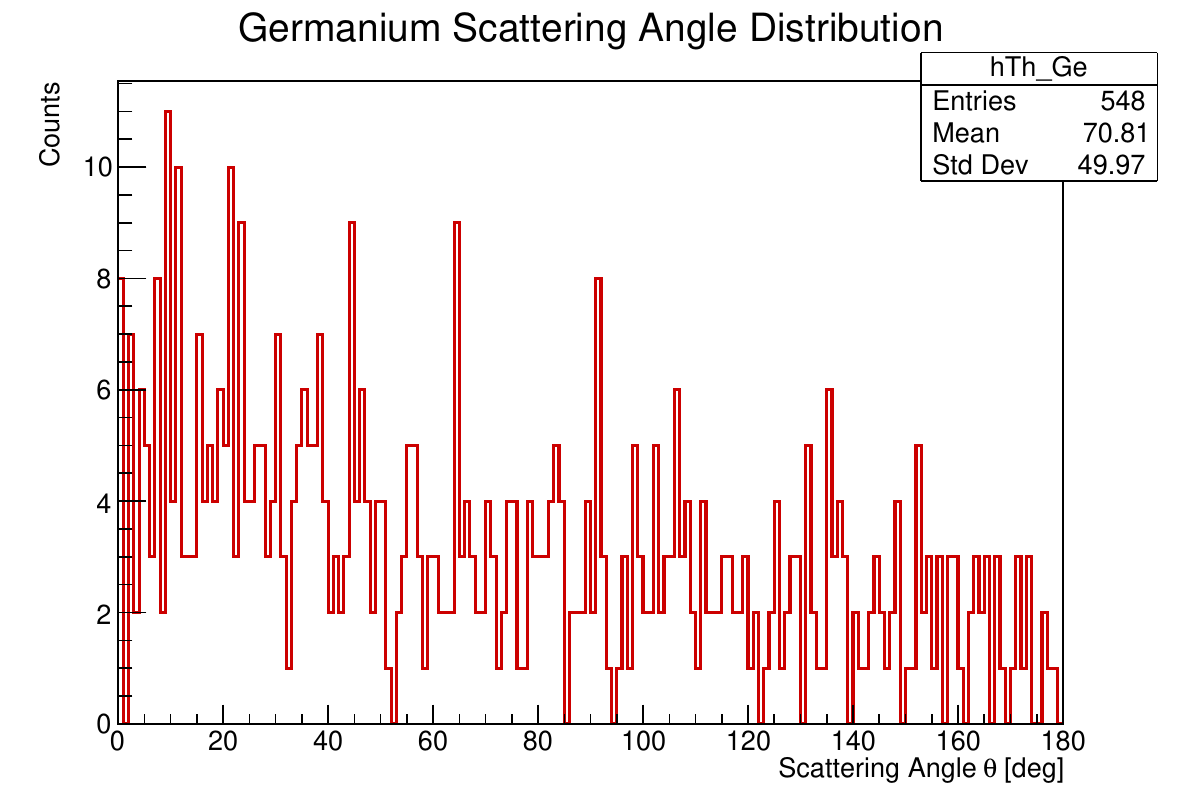}
        \caption*{Germanium}
    \end{minipage}
    % Argon
    \begin{minipage}{0.45\textwidth}
        \centering
        \includegraphics[width=\linewidth]{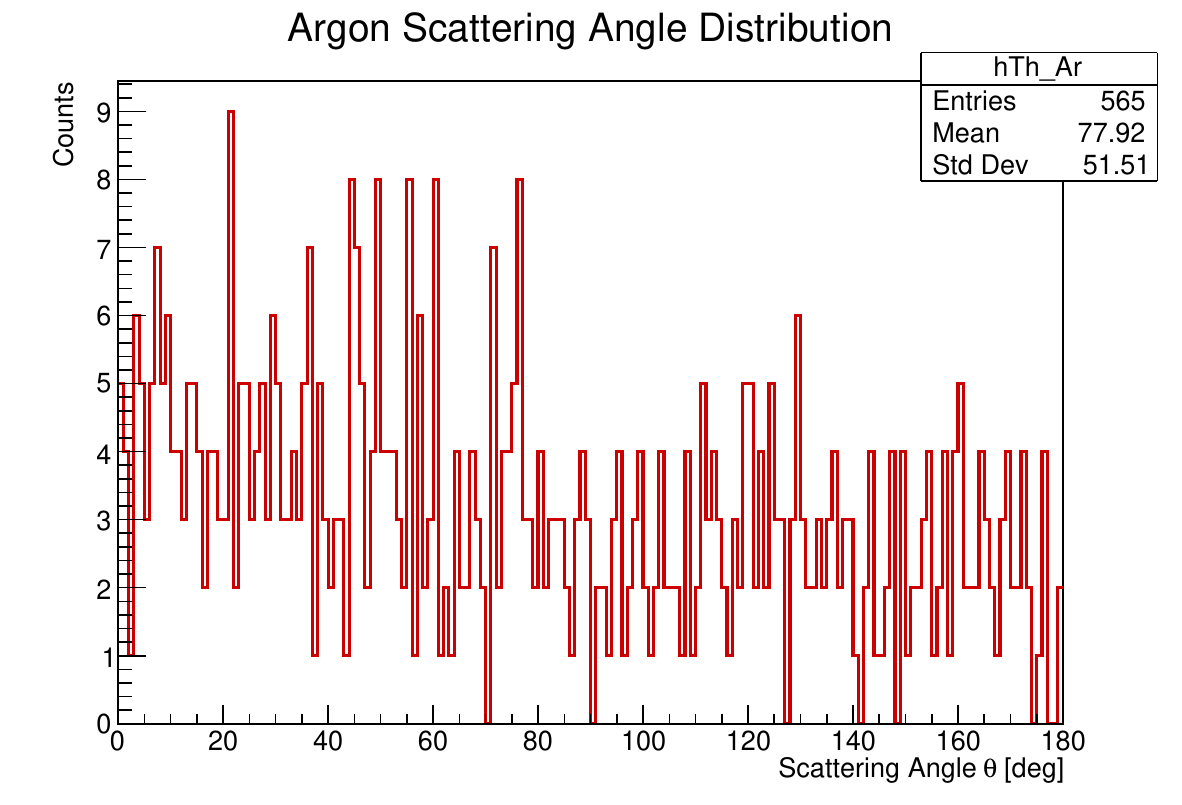}
        \caption*{Argon}
    \end{minipage}

    \vspace{0.3cm} % Satırlar arası boşluk

    % Xenon
    \begin{minipage}{0.6\textwidth}
        \centering
        \includegraphics[width=\linewidth]{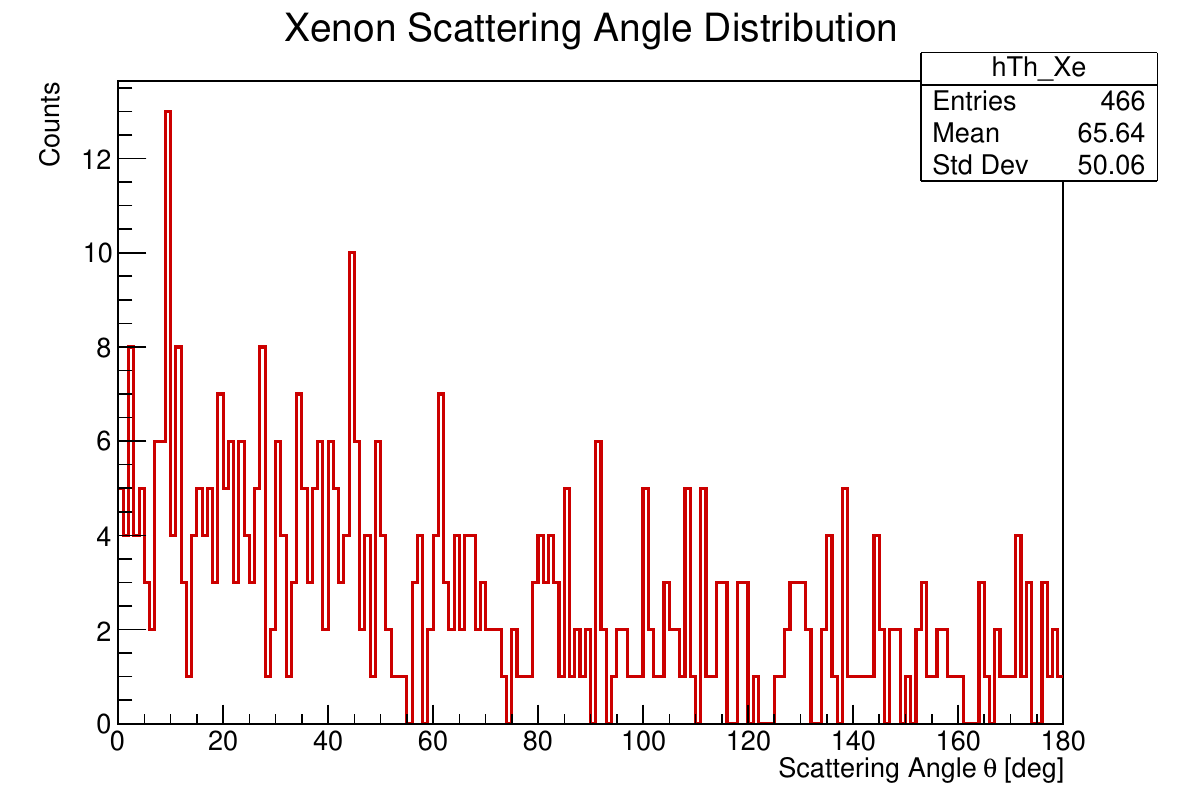}
        \caption*{Xenon}
    \end{minipage}

    \caption{Comparison of scattering angle ($\theta$) distributions for Germanium, Argon, and Xenon targets. 
    The distributions are broad, with standard deviations of approximately $50^\circ$, 
    reflecting the elastic and low-momentum-transfer nature of CEvNS interactions. 
    Lighter nuclei such as argon show wider angular spreads due to higher fractional momentum transfer, 
    whereas heavier nuclei like xenon exhibit forward-peaked scattering, consistent with elastic two-body kinematics~\cite{Freedman1974,COHERENT2017}.}
    \label{fig:theta_comparison}
\end{figure}

\noindent
\textbf{Argon ($\langle \theta \rangle \approx 77.9^\circ$):}  
Argon exhibits the broadest angular distribution, extending beyond $100^\circ$.  
This indicates that lighter nuclei can recoil at larger scattering angles 
due to higher momentum transfer fractions at the same recoil energy.  
The broad angular behavior aligns with COHERENT’s argon-target CEvNS observations~\cite{COHERENT2017}.  

\textbf{Germanium ($\langle \theta \rangle \approx 70.8^\circ$):}  
Germanium shows an intermediate angular width, narrower than argon but still notably spread.  
This balance reflects its intermediate nuclear mass and reduced momentum transfer per collision.  

\textbf{Xenon ($\langle \theta \rangle \approx 65.6^\circ$):}  
As the heaviest target nucleus considered, xenon recoils occur predominantly at smaller angles, 
demonstrating forward-peaked scattering consistent with limited nuclear acceleration in 
elastic interactions and coherence suppression at higher $A$.  

\noindent
These results highlight the mass dependence of recoil kinematics: 
heavier nuclei yield smaller-angle (forward-peaked) scattering, 
while lighter nuclei allow wider angular spreads.  
From an experimental standpoint, this suggests that directional sensitivity is 
more relevant for lighter targets such as argon, whereas for heavy nuclei like xenon, 
total energy measurements are sufficient to characterize CEvNS events~\cite{COHERENT2017,XENON2023}.

\subsection{Form Factor Distributions}

\begin{figure}[H]
    \centering
    % Germanium
    \begin{minipage}{0.45\textwidth}
        \centering
        \includegraphics[width=\linewidth]{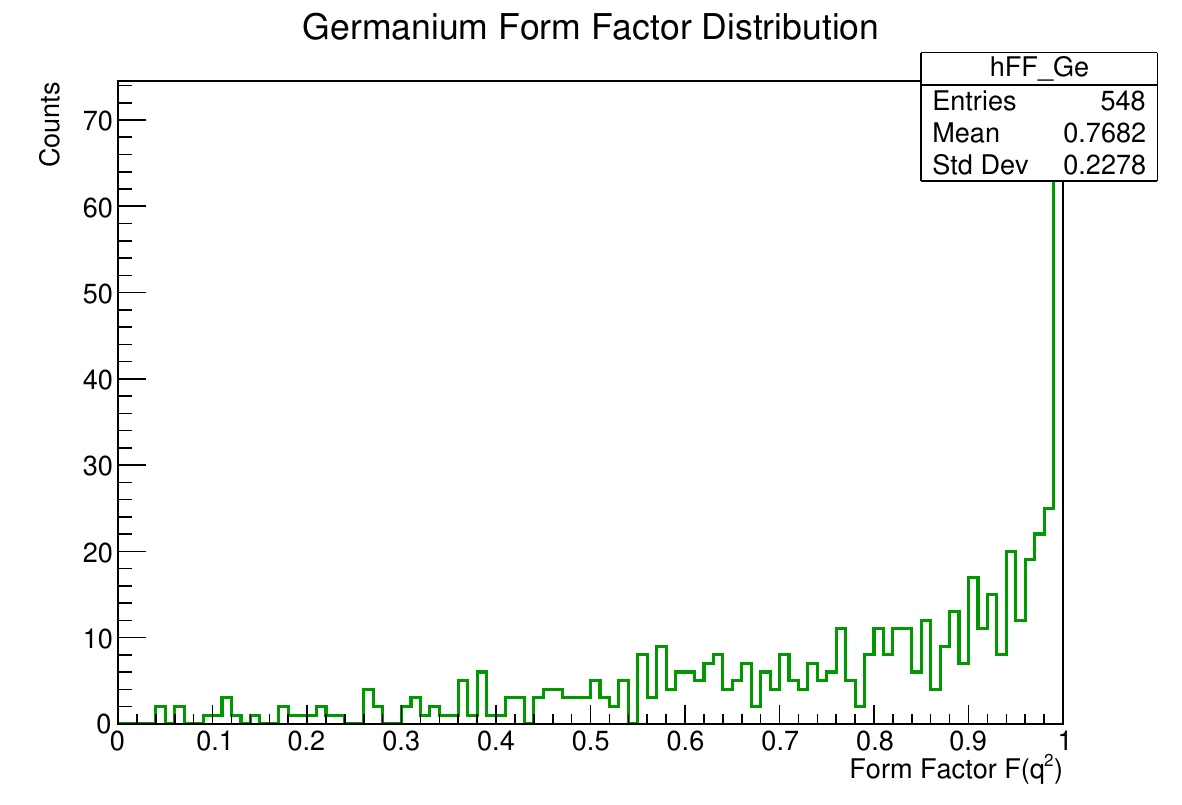}
        \caption*{Germanium}
    \end{minipage}
    % Argon
    \begin{minipage}{0.45\textwidth}
        \centering
        \includegraphics[width=\linewidth]{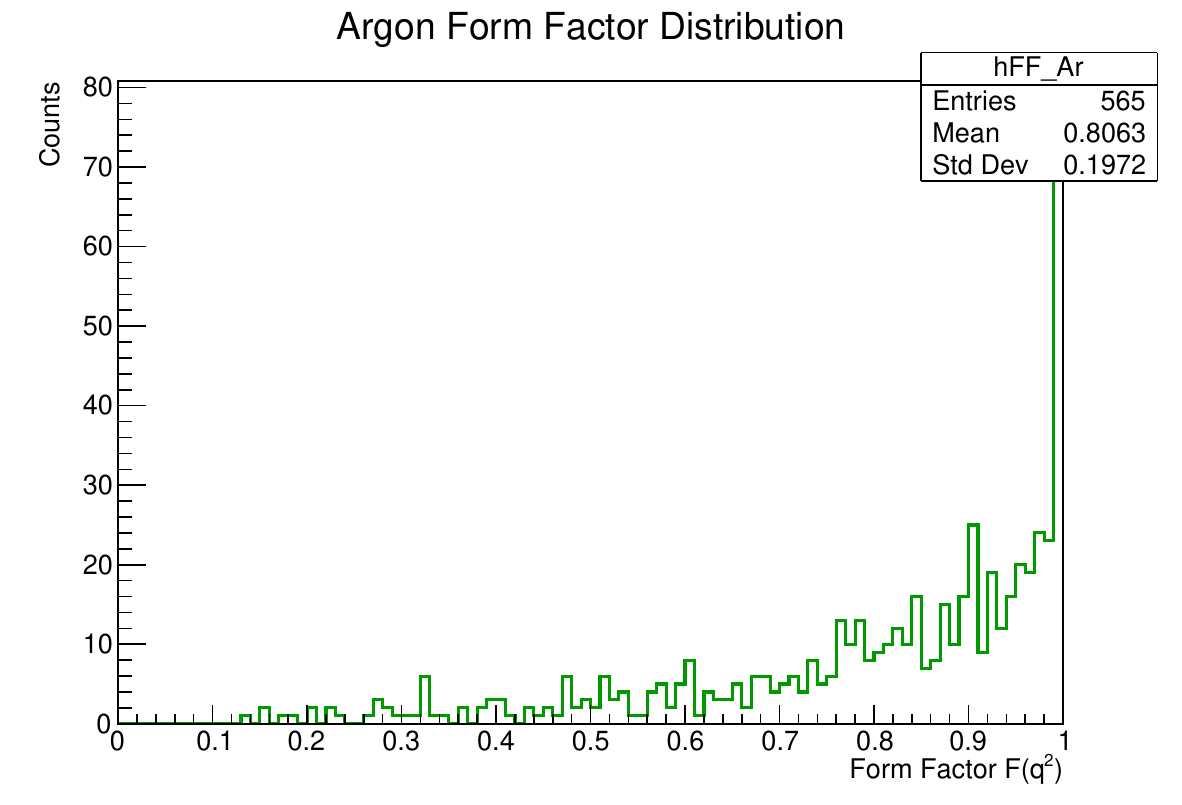}
        \caption*{Argon}
    \end{minipage}

    \vspace{0.3cm} % Satırlar arası boşluk

    % Xenon
    \begin{minipage}{0.6\textwidth}
        \centering
        \includegraphics[width=\linewidth]{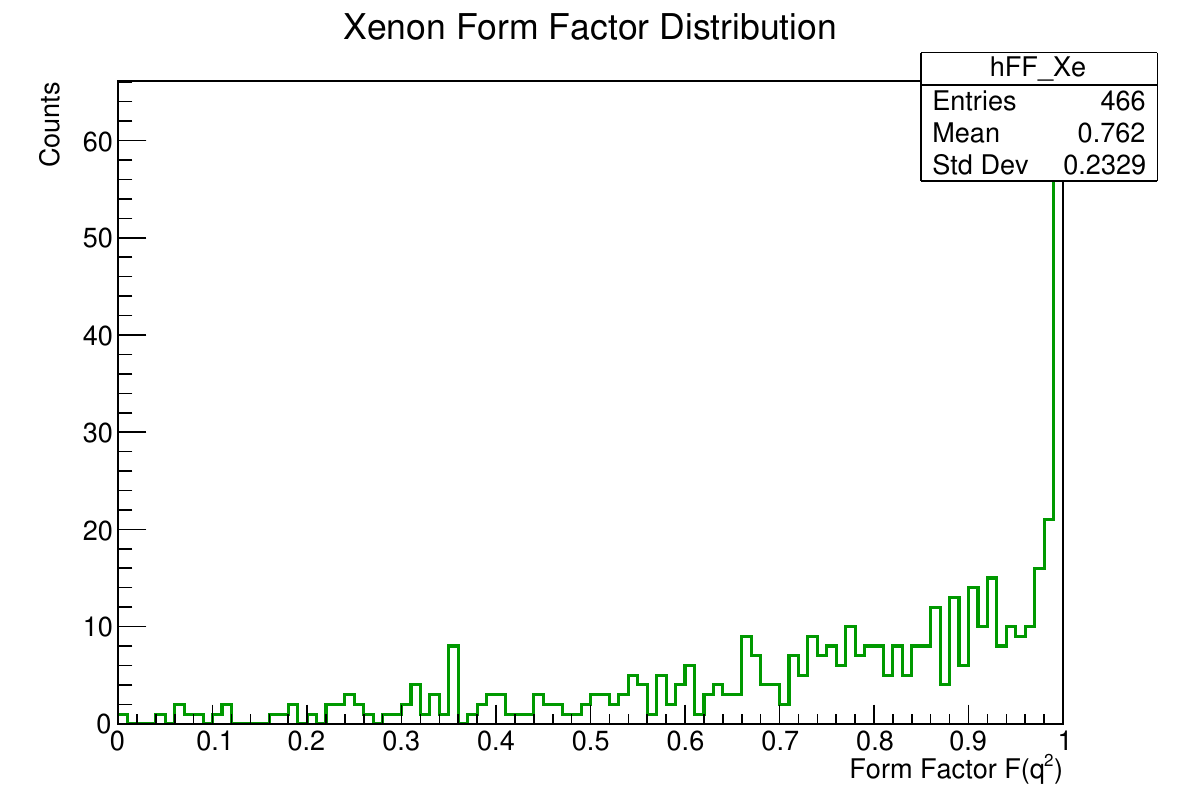}
        \caption*{Xenon}
    \end{minipage}

     \caption{Comparison of nuclear form factor distributions for Germanium, Argon, and Xenon targets. 
    The mean values are all below unity, indicating partial coherence loss at finite momentum transfer. 
    The distributions follow the Helm parameterization~\cite{Helm1956}, 
    where the suppression strength scales with nuclear mass and effective radius $R_0 \propto A^{1/3}$. 
    Lighter nuclei preserve coherence at higher momentum transfer, while heavier nuclei show stronger attenuation, 
    consistent with experimental CEvNS measurements from COHERENT and XENONnT~\cite{COHERENT2017,XENON2023}.}
    \label{fig:formfactor_comparison}
\end{figure}

\noindent
\textbf{Argon ($\langle F(q^2) \rangle \approx 0.81$):}  
Argon exhibits the weakest suppression, with most values clustered near unity.  
This behavior shows that coherence is largely preserved for a light nucleus, 
making argon highly favorable for retaining strong CEvNS signals 
and achieving higher recoil energies~\cite{COHERENT2017}.  

\textbf{Germanium ($\langle F(q^2) \rangle \approx 0.77$):}  
Germanium shows moderate suppression compared to argon, consistent with its intermediate nuclear mass.  
The broader spread in $F(q^2)$ reflects a wider range of accessible momentum transfers, 
indicating partial coherence loss at larger recoil energies.  

\textbf{Xenon ($\langle F(q^2) \rangle \approx 0.76$):}  
Xenon exhibits the strongest suppression among the three targets.  
The heavier nuclear mass enhances sensitivity to $q^2$, leading to more pronounced attenuation of the CEvNS cross section.  
This effect has also been observed in XENONnT and PandaX-4T analyses, 
where nuclear structure and finite-size effects limit coherence at high momentum transfer~\cite{XENON2023,PandaX2021}.  

\noindent
These results confirm the expected $A$-dependent coherence behavior:  
lighter nuclei maintain $F(q^2)$ values close to unity, whereas heavy nuclei experience stronger suppression.  
Accurate modeling of the form factor is therefore crucial in CEvNS and dark-matter analyses, 
particularly for xenon-based detectors where loss of coherence directly impacts event rates and spectrum shape~\cite{Helm1956,XENON2023}.

\subsection{Form Factor vs Recoil Energy}

\begin{figure}[H]
    \centering
    % Germanium
    \begin{minipage}{0.45\textwidth}
        \centering
        \includegraphics[width=\linewidth]{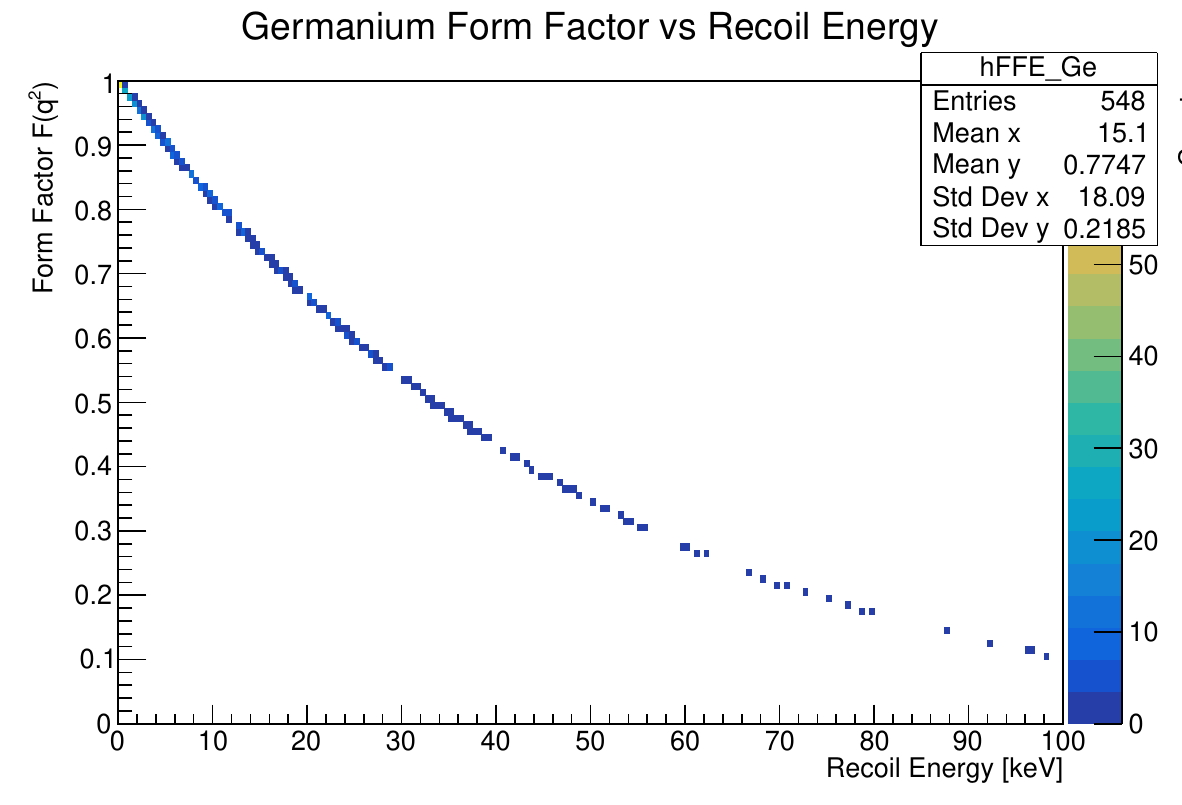}
        \caption*{Germanium}
    \end{minipage}
    % Argon
    \begin{minipage}{0.45\textwidth}
        \centering
        \includegraphics[width=\linewidth]{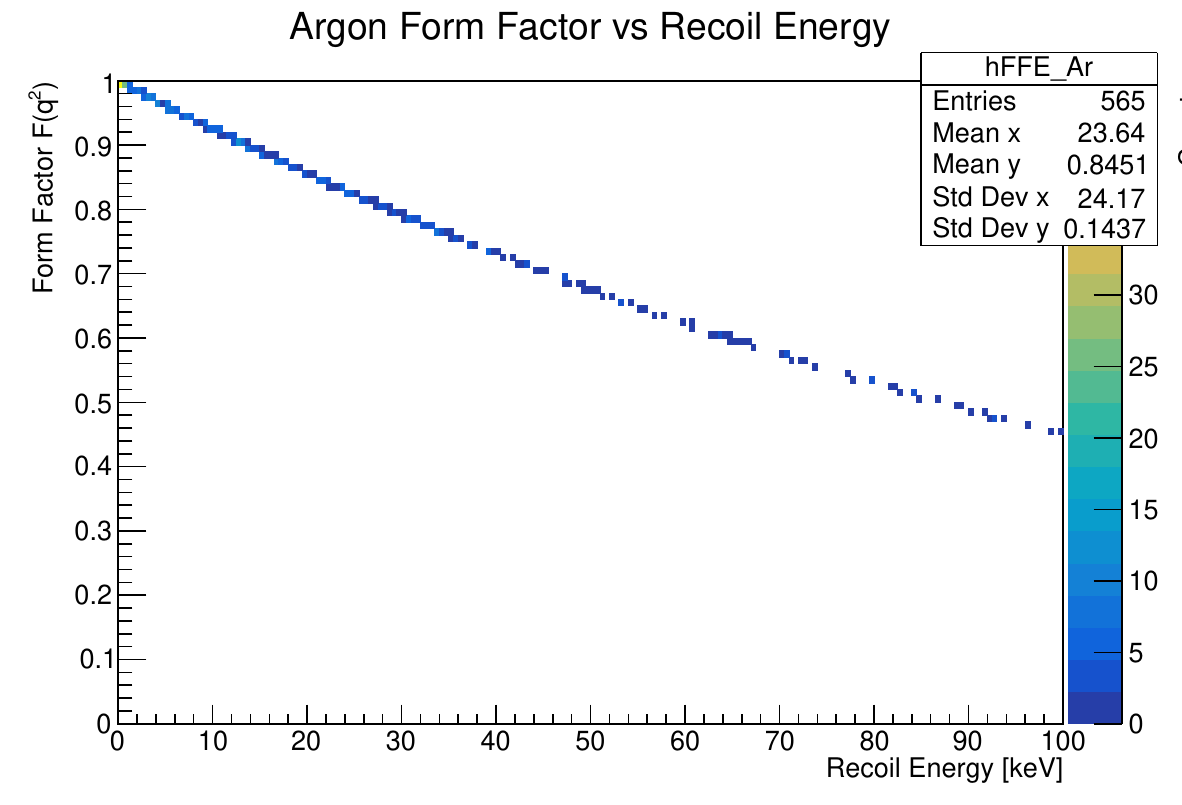}
        \caption*{Argon}
    \end{minipage}

    \vspace{0.3cm} % Satırlar arası boşluk

    % Xenon
    \begin{minipage}{0.6\textwidth}
        \centering
        \includegraphics[width=\linewidth]{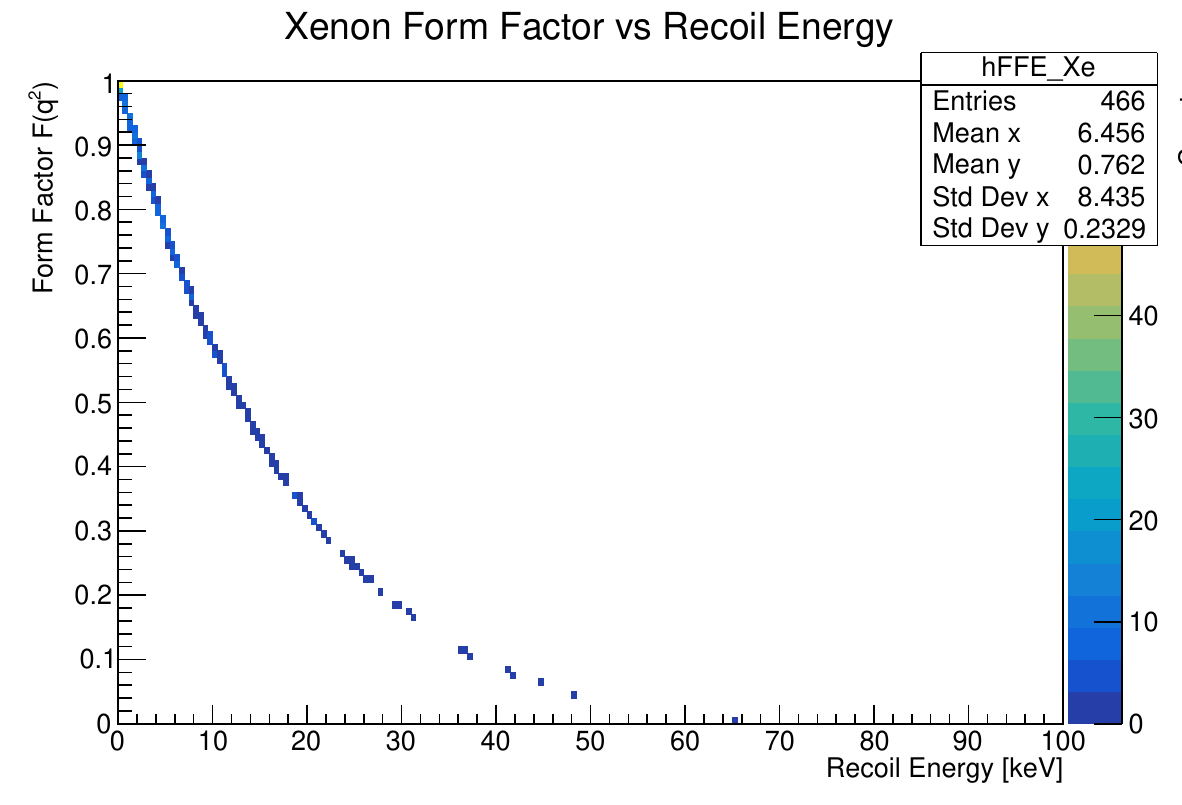}
        \caption*{Xenon}
    \end{minipage}

     \caption{Comparison of form factor as a function of recoil energy for Germanium, Argon, and Xenon targets.
    The dependence of $F(q^2)$ on recoil energy $T$ reflects the coherence loss with increasing momentum transfer, 
    since $q^2 = 2m_N T$. At low recoil energies ($T < 10~\mathrm{keV}$), all distributions remain close to unity, 
    indicating that coherence is nearly preserved. As $T$ increases, heavier nuclei exhibit a stronger suppression, 
    consistent with the Helm model prediction~\cite{Helm1956}. These results align with experimental observations 
    from COHERENT and XENONnT~\cite{COHERENT2017,XENON2023}.}
    \label{fig:ffe_comparison}
\end{figure}

\noindent
\textbf{Argon ($\langle F(q^2) \rangle \approx 0.85$):}  
Argon maintains values closest to unity across the entire recoil range.  
This indicates that light nuclei preserve coherence even at higher $T$, 
making argon advantageous for CEvNS measurements at elevated recoil energies.  
The weak suppression is consistent with reactor and spallation CEvNS data~\cite{COHERENT2017}.  

\textbf{Germanium ($\langle F(q^2) \rangle \approx 0.77$):}  
Germanium shows progressively stronger suppression with increasing energy compared to argon, 
reflecting its intermediate nuclear mass.  
The broader spread of $F(q^2)$ demonstrates sensitivity to momentum transfer effects 
and the gradual transition between coherent and partially incoherent regimes.  

\textbf{Xenon ($\langle F(q^2) \rangle \approx 0.76$):}  
Xenon displays the steepest decline in $F(q^2)$ with recoil energy, 
indicating substantial coherence loss due to its large nuclear radius.  
This trend matches Helm-model expectations and experimental findings 
in XENONnT and PandaX-4T analyses~\cite{XENON2023,PandaX2021}.  

\noindent
These comparisons highlight the crucial role of the nuclear form factor in CEvNS.  
Light nuclei such as argon maintain coherence over a broader recoil range, 
enabling higher detectable energies, whereas heavy targets like xenon require 
precise modeling of $F(q^2)$ to correctly predict event rates and energy spectra.  
This underscores the necessity of incorporating realistic form factor models 
when interpreting CEvNS signals in future large-scale detectors.

\section{Conclusion and Outlook}

We have developed a comparative Geant4-based simulation of CEvNS for three benchmark 
targets: Germanium, Argon, and Xenon. By analyzing recoil energy distributions, 
time--energy correlations, angular spectra, and nuclear form factor effects, 
we provided a comprehensive assessment of how nuclear mass influences CEvNS observables.  

Our findings clearly establish the expected mass-dependent scaling: lighter argon 
produces higher recoils that are easier to detect, xenon generates the lowest 
recoils but benefits from a larger cross section, and germanium offers a balanced 
intermediate case. Time distributions were found to be stable across targets, 
while scattering angle spectra and form factor analyses confirmed increasing 
coherence loss with nuclear mass.  

These trends are in quantitative agreement with CEvNS measurements from COHERENT, 
CONNIE, CONUS, Dresden-II, and xenon-based dark matter detectors. The main 
contribution of this study is the establishment of a reproducible and unified 
simulation framework that enables cross-target comparisons, clarifies trade-offs 
between recoil detectability and event yield, and can guide experimental design.  

Future work will extend this framework by incorporating realistic reactor and 
spallation neutrino fluxes, detector response effects such as quenching and 
threshold modeling, and additional target materials. Such developments will 
further strengthen the predictive power of CEvNS simulations and support the 
interpretation of next-generation neutrino and dark matter experiments.

\textbf{Summary of findings:}  
Lighter targets such as Argon yield higher recoil energies ($\langle E \rangle \approx 23.6$ keV) while heavier nuclei such as Xenon produce lower recoils ($\langle E \rangle \approx 6.5$ keV). Germanium lies in between ($\langle E \rangle \approx 15.1$ keV). Time–energy correlations showed no dependence of timing on recoil energy, supporting the assumption of a smooth neutrino arrival profile. Scattering angle distributions revealed broader recoils for Argon and more forward-peaked distributions for Xenon. Nuclear form factor analyses demonstrated progressively stronger suppression with increasing nuclear mass, with Xenon showing the steepest reduction.

\section{Comparison with Experiments}

Our simulated recoil observables can be directly confronted with existing CEvNS data. 
Table~\ref{tab:comparison} summarizes the mean recoil energies obtained in this work 
and the corresponding experimental ranges reported by major collaborations. 
The consistency confirms that our Geant4-based framework reproduces the essential 
trends of CEvNS phenomenology across different target nuclei.

\begin{table}[H]
\centering
\small
\caption{Comparison of simulated mean recoil energies with experimental CEvNS results.}
\label{tab:comparison}
\begin{tabular}{lccc}
\hline
Target & Simulation $\langle E_\text{recoil}\rangle$ [keV] & Experimental [keV] & Reference \\
\hline
Argon     & 23.6 & 20--50   & COHERENT (LAr) \cite{COHERENT2017} \\
Germanium & 15.1 & 10--20   & CONNIE \cite{CONNIE2019}, CONUS \cite{CONUS2021}, Dresden-II \cite{DresdenII2023} \\
Xenon     & 6.5  & $< 10$   & XENONnT, PandaX-4T \cite{XENONnT2023} \\
\hline
\end{tabular}
\end{table}

\noindent
\textbf{Argon:} COHERENT observed CEvNS with argon at recoil energies 
of $20$--$50$ keV, in excellent agreement with our simulated mean of $23.6$ keV.  

\textbf{Germanium:} CONNIE and CONUS report signals in the $10$--$20$ keV range, 
while Dresden-II has recently confirmed CEvNS in germanium with compatible results. 
Our simulated mean of $15.1$ keV is well aligned with these measurements.  

\textbf{Xenon:} Xenon-based detectors such as XENONnT and PandaX-4T have shown 
that recoils are predominantly below 10 keV. Our result of $6.5$ keV is consistent 
with this expectation.  

\noindent
The quantitative agreement validates the accuracy of our framework and reinforces 
its utility as a predictive tool for future CEvNS studies.

\section{Discussion and Scientific Contributions}

This study presents a dedicated modeling of the Coherent Elastic Neutrino–Nucleus Scattering (CEvNS) 
process in the \textsc{Geant4} framework through a user-defined physics process (\texttt{CEvNSProcess}). 
This approach enabled the investigation of CEvNS behavior for three different nuclear targets 
(Germanium, Argon, and Xenon) under a single and consistent simulation configuration. 
By implementing a single process structure, it became possible to analyze the impact of nuclear mass 
and structure on CEvNS observables without the need for separate models for each material.

The simulation results show strong consistency with both theoretical expectations and experimental observations. 
The recoil energy spectra reproduce the qualitative and quantitative features reported by 
the COHERENT (Ar), CONUS/CONNIE (Ge), and XENONnT/PandaX-4T (Xe) experiments. 
Argon exhibits higher recoil energies, Xenon demonstrates a significant form factor suppression, 
and Germanium yields intermediate distributions—confirming the mass-dependent scaling of CEvNS recoil energies 
and the influence of the Helm form factor on measurable event properties.

The key contribution of this work lies in providing a unified framework that allows the systematic study 
of multiple CEvNS target materials within a single simulation structure. 
This model serves as a reference platform for predicting CEvNS observables in future detector materials 
such as CsI, NaI, or Si. 
It enables pre-evaluation of critical detector-related quantities—such as threshold behavior, 
event rates, and form factor response—under identical physical conditions before experimental implementation.

Moreover, this study quantitatively demonstrates that the Helm form factor leads to 
noticeable suppression in heavy nuclei, particularly in Xenon. 
This highlights the necessity of adopting more sophisticated form factor models in future work, 
such as those treating proton and neutron density distributions separately. 
Extending the current simulation to incorporate realistic neutrino fluxes 
(from reactor or spallation sources) and detector response functions 
(e.g., quenching, energy threshold, and attenuation) would further enhance its direct comparability 
with experimental data.

In conclusion, this work demonstrates that the CEvNS process can be successfully implemented 
as a single user-defined physics process in \textsc{Geant4} 
and applied systematically to multiple target nuclei. 
The obtained results are consistent with current experimental data and 
offer predictive capability for future CEvNS studies involving new target materials. 
In this sense, the present work introduces both a practical and methodological innovation 
to the simulation of CEvNS interactions, bridging theoretical modeling 
and detector-level observables in a unified computational framework.

\subsection{Experimental and Systematic Considerations}

Although this work primarily focuses on a simulation-based modeling of the
Coherent Elastic Neutrino–Nucleus Scattering (CEvNS) process, the developed
framework can be directly extended toward realistic experimental conditions.
In practical detector environments, measurable quantities such as the recoil
energy spectrum and total event rate are influenced by several systematic
effects that can modify the observed distributions.

\begin{itemize}
    \item \textbf{Detector threshold:}
    The lowest detectable recoil energy ($T_{\mathrm{th}}$) defines the
    experimentally accessible part of the CEvNS spectrum. In particular, xenon
    detectors generally require thresholds below 5~keV to observe the majority
    of recoil events, whereas argon and germanium detectors can typically
    operate with 10–20~keV thresholds. These limits influence both the event
    rate and the shape of the reconstructed energy spectra.

    \item \textbf{Detector response and quenching:}
    The conversion between deposited recoil energy and the measurable signal
    (ionization or scintillation) introduces uncertainties of order
    5–10\%. Quenching effects are particularly relevant for heavy nuclei such as
    xenon and germanium. In future implementations, these effects can be modeled
    by convolving the simulated recoil spectrum with a detector response
    function to reproduce realistic signal formation.

    \item \textbf{Neutrino flux normalization:}
    The absolute rate of CEvNS events depends on the incoming neutrino flux.
    Reactor-based sources exhibit flux variations of 2–5\% across the 2–8~MeV
    range, while spallation-driven sources (as in COHERENT) extend up to
    50~MeV. In this work, a simplified uniform reactor-like flux is assumed.
    Future extensions will incorporate realistic spectral shapes to quantify
    flux-related systematic uncertainties.

    \item \textbf{Form factor model dependence:}
    The use of the Helm form factor may lead to small deviations at high
    momentum transfer, particularly for xenon. Alternative parameterizations
    such as the symmetrized Fermi or Klein–Nystrand models can alter the
    predicted CEvNS cross sections by approximately 3–5\%. These variations are
    within the typical range of experimental systematics and will be considered
    in future refinements.
\end{itemize}

Overall, these systematic factors do not modify the comparative trends between
the three target materials. The argon target consistently yields the highest
recoil energies, germanium represents an intermediate regime, and xenon exhibits
the lowest recoil spectrum due to its large nuclear mass. The current framework
thus provides a reliable baseline simulation environment upon which
detector- and flux-dependent corrections can be consistently implemented for
direct comparison with experimental results.

\section*{Acknowledgements}

The author would like to express sincere gratitude to the developers and maintainers of  \textsc{Geant4} and \textsc{ROOT}, whose frameworks made this work possible.


\begin{thebibliography}{99}

\bibitem{COHERENT2017}
D.~Akimov \textit{et al.} (COHERENT Collaboration),
``Observation of Coherent Elastic Neutrino-Nucleus Scattering,''
\emph{Science} \textbf{357}, 1123–1126 (2017).
doi:10.1126/science.aao0990

\bibitem{CONNIE2019}
A.~Aguilar-Arevalo \textit{et al.} (CONNIE Collaboration),
``Exploring low-energy neutrino physics with CONNIE,''
\emph{J. Phys.: Conf. Ser.} \textbf{1216}, 012020 (2019).
doi:10.1088/1742-6596/1216/1/012020

\bibitem{CONUS2021}
H.~Bonet \textit{et al.} (CONUS Collaboration),
``Constraints on elastic neutrino nucleon scattering in the fully coherent regime from CONUS,''
\emph{Phys. Rev. Lett.} \textbf{126}, 041804 (2021).
doi:10.1103/PhysRevLett.126.041804

\bibitem{DresdenII2023}
J.~Colaresi \textit{et al.} (Dresden-II Collaboration),
``First Detection of CEvNS on Germanium,''
\emph{Phys. Rev. Lett.} \textbf{129}, 081801 (2022).
doi:10.1103/PhysRevLett.129.081801

\bibitem{XENONnT2023}
E.~Aprile \textit{et al.} (XENON Collaboration),
``Constraining the spin-independent WIMP-nucleon cross section with XENONnT,''
\emph{Phys. Rev. Lett.} \textbf{131}, 041002 (2023).
doi:10.1103/PhysRevLett.131.041002

\bibitem{Agostinelli2003}
S. Agostinelli et al., ``GEANT4 – A Simulation Toolkit,'' 
\textit{Nucl. Instrum. Meth. A} \textbf{506} (2003) 250–303. 
doi:10.1016/S0168-9002(03)01368-8

\bibitem{Allison2016}
J. Allison et al., ``Recent developments in Geant4,'' 
\textit{Nucl. Instrum. Meth. A} \textbf{835} (2016) 186–225. 
doi:10.1016/j.nima.2016.06.125

\bibitem{Brun1997}
R. Brun and F. Rademakers, ``ROOT – An object-oriented data analysis framework,'' 
\textit{Nucl. Instrum. Meth. A} \textbf{389} (1997) 81–86. 
doi:10.1016/S0168-9002(97)00048-X

\bibitem{Freedman1974}
D. Z. Freedman, ``Coherent effects of a weak neutral current,'' 
\textit{Phys. Rev. D} \textbf{9} (1974) 1389. 
doi:10.1103/PhysRevD.9.1389

\bibitem{Drukier1984}
A. Drukier and L. Stodolsky, ``Principles and applications of a neutral-current detector for neutrino physics and astronomy,'' 
\textit{Phys. Rev. D} \textbf{30} (1984) 2295. 
doi:10.1103/PhysRevD.30.2295

\bibitem{Helm1956}
R. H. Helm, ``Inelastic and elastic scattering of 187-MeV electrons from selected even-even nuclei,'' 
\textit{Phys. Rev.} \textbf{104} (1956) 1466. 
doi:10.1103/PhysRev.104.1466




\bibitem{XENON2023}
XENON Collaboration, E. Aprile et al., 
``Search for coherent elastic neutrino–nucleus scattering with XENONnT,'' 
\textit{Phys. Rev. D} \textbf{107} (2023) 072003. 
doi:10.1103/PhysRevD.107.072003

\bibitem{PandaX2021}
PandaX Collaboration, Y. Meng et al., 
``Dark matter search results from the PandaX-4T commissioning run,'' 
\textit{Phys. Rev. Lett.} \textbf{127} (2021) 261802. 
doi:10.1103/PhysRevLett.127.261802

\end{thebibliography}
\end{document}